\documentclass[aps,letterpaper,prb,twocolumn,notitlepage,showpacs,nobalancelastpage,
    amsmath]{revtex4-1}
\usepackage{bbm,color,amsmath,amssymb,graphicx}
\usepackage[pdftex]{hyperref}
\usepackage{braket}
\usepackage{stmaryrd}

\providecommand{\average}[1]{\left\llbracket{#1}\right\rrbracket}

\begin{document}

\title{Intrinsic decoherence in isolated quantum systems}
\date{\today}

\author{Yang-Le Wu}
\author{Dong-Ling Deng}
\author{Xiaopeng Li}
\author{S. Das Sarma}
\affiliation{Condensed Matter Theory Center and Joint Quantum Institute, Department
of Physics, University of Maryland, College Park, Maryland 20742-4111, USA}

\begin{abstract}
We study the intrinsic, disorder-induced decoherence of an isolated quantum 
system under its own dynamics.
Specifically, we investigate the characteristic time scale (i.e., the 
decoherence time) associated with an interacting many-body system losing the 
memory of its initial state.
To characterize the erasure of the initial state memory, we define a 
time scale, the intrinsic decoherence time, by thresholding the 
gradual decay of the disorder-averaged return probability.
We demonstrate the system-size independence of the intrinsic decoherence time 
in different models,
and we study its dependence on the disorder strength.
We find that the intrinsic decoherence time increases monotonically as the 
disorder strength increases in accordance with the relaxation of locally 
measurable quantities.
We investigate several interacting spin (e.g., Ising and Heisenberg) and 
fermion (e.g., Anderson and Aubry-Andr\'e) models to obtain the intrinsic 
decoherence time as a function of disorder and interaction strength.
\end{abstract}

\maketitle

\section{Introduction}

Decoherence is an important issue in the practical realization of quantum 
computers, as it hinders the preservation and controlled manipulation of 
quantum information~\cite{DiVincenzo00:QC}.
Combating decoherence in various quantum computing architectures has thus 
attracted a lot of attention from both theoretical and experimental 
communities in recent years.
In this line of research, decoherence typically refers to the loss of quantum 
coherence in a system of controlled qubits due to the entanglement generated 
by inevitable couplings to the external environment.
Significant efforts and progress have been made to reduce such couplings while 
maintaining the ability to reliably control the quantum 
system~\cite{Ladd10:Review,Zwanenburg13:QE,Devoret13:SC}.

In this paper we investigate a related, but distinct concept, the ``intrinsic 
decoherence'' in an \emph{isolated} quantum system without any explicit 
coupling to the external environment.
The decoherence here refers to the erasure of local quantum information in an 
initial non-eigenstate as the isolated system relaxes by itself (through 
interaction) to thermalization.
This is motivated by recent developments in the study of many-body 
localization~\cite{Basko06:MBL,Oganesyan07:MBL,Pal10:MBL,Imbrie14:Proof,
Schreiber15:MBL,Smith15:MBL,Choi16:MBL,Bordia16:MBL},
where an interacting, isolated quantum system resists 
thermalization under strong disorder and retains indefinitely the quantum 
memory in its initial, highly excited 
state~\cite{Altman15:MBLReview,Nandkishore15:MBLReview}.
Here we take a closer look at the thermalization (or relaxation) dynamics in 
terms of both local observables and the global return probability.
We aim to construct a quantitative measure of the gradual erasure of the 
initial state information during the time evolution of the isolated system
and study the dependence of the erasure rate on the disorder strength.
The decoherence considered here arises from an intricate interplay between 
external noise (as manifested in the disorder) and interaction among all the 
qubits in the circuit, and is thus quite distinct from the corresponding 
purely environmental single-qubit decoherence considered in the context of 
quantum computing architectures.  We use the word ``decoherence'' here in a 
loose qualitative manner (for example, ``relaxation'' rather than 
``decoherence'' would have been an equally appropriate nomenclature here) but 
define the intrinsic decoherence time rather precisely (with respect to the 
initial state) so that there is no semantic confusion. 
The important point to be emphasized is that the intrinsic decoherence being 
studied here is a property of the \emph{whole} interacting many-body 
system, and is not a single-qubit phenomenon.

We define a quantity, dubbed the ``intrinsic decoherence time,'' by 
thresholding the decay of the disorder-averaged return probability after 
a certain rescaling.
This quantity captures the time scale over which the system loses track of the 
local information in the initial state under its own dynamics.
We pay special attention to ensure that the intrinsic decoherence time thus 
defined is insensitive to finite-size effects and extrapolates properly to the 
thermodynamic limit.

We study this quantity using exact diagonalization in four different 
models for both spins and interacting fermions.
There is background disorder in all the models, and the models are, by 
definition, interacting models as each spin or fermion interacts with others.
Across the different models, we consistently find that the intrinsic 
decoherence time increases monotonically as the disorder strength increases 
(presumably diverging in the many-body-localized systems where local initial 
state memory is preserved forever),
and it only displays rather moderate dependence on system size with a 
reasonable scaling behavior.
For comparison, we have also computed similar time scales extracted from 
local observables such as magnetization and density imbalance.
We find that these time scales also exhibit similar monotonic dependence 
on disorder, giving confidence in the belief that the intrinsic decoherence 
time is an experimentally accessible physical property characterizing the 
relaxation dynamics of disordered interacting many-body systems.

The paper is organized as follows.
In Sec.~\ref{sec:Ising} we examine the dynamics of the disorder-averaged 
return probability in the transverse-field Ising model and discuss the proper 
definition of the intrinsic decoherence time based on the per-site return 
probability.
We further characterize the intrinsic decoherence time in another three
different models, namely,
the nearest-neighbor Heisenberg model with a random magnetic field
in Sec.~\ref{sec:Heisenberg},
the Anderson model of the interacting fermions with a random on-site 
potential in Sec.~\ref{sec:Anderson},
and also the interacting Aubry-Andr\'e incommensurate model
in Sec.~\ref{sec:AA}.
We conclude the paper in Sec.~\ref{sec:conclusion}.

\section{Intrinsic decoherence in transverse-field Ising model}
\label{sec:Ising}

We aim to characterize the intrinsic decoherence in an isolated, disordered 
quantum system.
As a specific example, we consider a linear array of coupled trapped-ion 
qubits, described by the disordered transverse-field Ising 
model~\cite{Porras04:Heff,Smith15:MBL}:
\begin{equation}
H_\text{Ising}=
\sum_{i<j}J_{i,j}\sigma_i^x\sigma_j^x
+\frac{1}{2}\sum_ih_i\sigma_i^z.
\end{equation}
With some minor modifications, the same Hamiltonian, in principle, applies 
also to the capacitively coupled, singlet-triplet semiconductor spin qubits.
The interqubit couplings take the power-law form $J_{i,j}={J_0/|i-j|^\alpha}$, 
while the transverse fields $h_i$ are drawn uniformly from ${[h_0-W,h_0+W]}$.
This model has recently been experimentally demonstrated to 
exhibit features of many-body localization for disorder $W$ stronger than a 
few $J_0$\cite{Smith15:MBL}.
In this paper, we work in the short-range coupled regime and pick $\alpha=3$ 
and $h_0=4J_0$ such that the finite-size localization crossover is relatively 
unambiguous.
In an earlier work~\cite{Wu16:Iontrap}, it was established that
$\alpha>1$ ($<1$) appears to separate qualitatively distinct regimes 
manifesting localization (no localization), leading to our choice of $\alpha=3$ 
in the current work so that the system is relatively deep inside the 
many-body-localized phase.

We use exact diagonalization to study the dynamics of the disordered system.
We prepare the coupled qubits in the N\'eel initial state 
$\ket{\psi(0)}=\ket{\uparrow\downarrow\cdots\uparrow\downarrow}$
and track their time evolution.
We define the disorder-averaged return probability to the initial state as
\begin{equation}\label{eq:R}
R(t)=\average{\big|\langle\psi(t)|\psi(0)\rangle\big|^2},
\end{equation}
where the double brackets $\average{\cdot}$ denote averaging over disorder 
realizations. For the results presented in this paper, we typically average 
over $10^3$ disorder samples for each parameter set.
Obviously, the dynamics of the return probability depends both on the 
Hamiltonian and the initial state (chosen to be the standard N\'eel state here).

\begin{figure}[]
\centering
\includegraphics{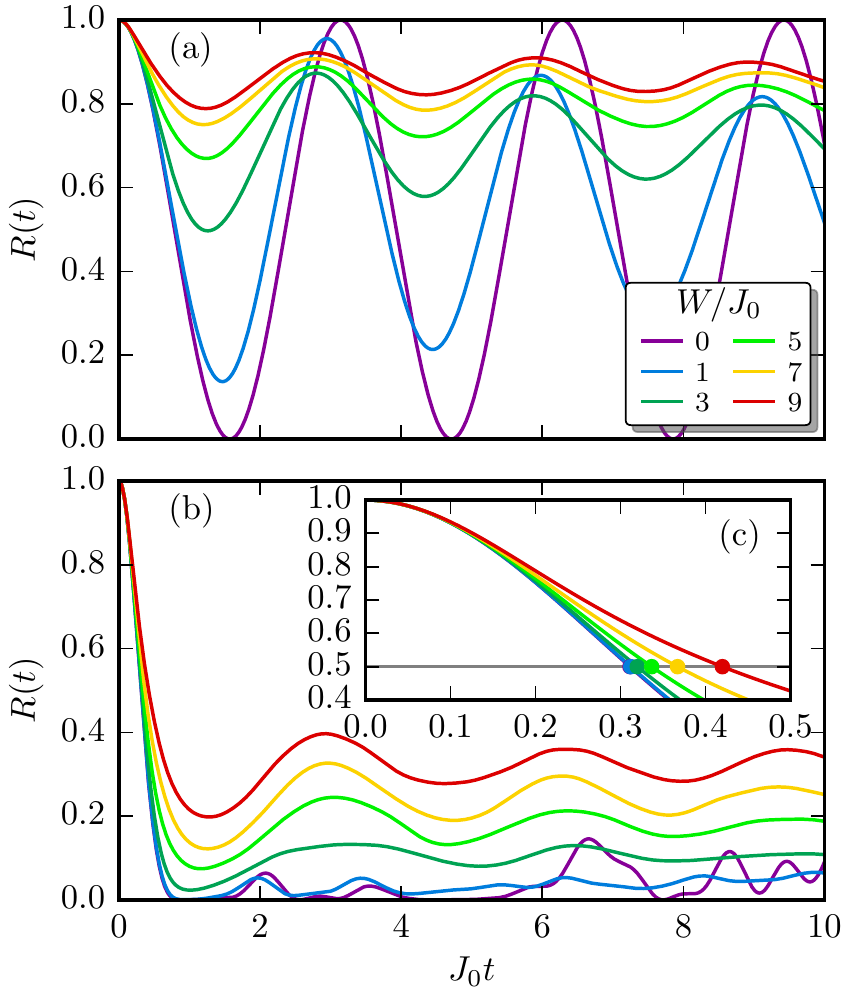}%
\caption{\label{fig:ising-rp-dynamics}
Dynamics of the disorder-averaged return probability $R(t)$ for (a) $N=2$ and 
(b) $N=8$ Ising qubits for various disorder strengths $W$.
The inset (c) illustrates the thresholding procedure that defines $T$.
The three panels share the same color code illustrated in (a).
}
\end{figure}

Figure~\ref{fig:ising-rp-dynamics} shows the time dependence of the 
disorder-averaged return probability $R(t)$ for $N=2$ and $8$ qubits under 
various strengths of disorder.
For $N=2$, we observe a persistent oscillation of $R(t)$ with a decaying 
envelope. Mathematically this is reminiscent of (although physically different 
from) the damped Rabi oscillations of a single qubit due to couplings to the 
external environment.
Naively, one may try to extract a decoherence time scale from the decaying 
envelope in the same spirit as the relaxation time $T_1$ obtained from 
Rabi oscillations~\cite{Zwanenburg13:QE}.
Unfortunately, this intuitive picture of a damped oscillator breaks down once 
the system grows beyond a couple of qubits, and no such simple analogy to Rabi 
oscillations or noninteracting single-qubit systems is possible in the 
multiqubit situation.
In Fig.~\ref{fig:ising-rp-dynamics}(b), we find that for $N=8$ qubits, the 
disorder-averaged return probability $R(t)$ does not exhibit a gradual decay 
of oscillations.
Instead, $R(t)$ undergoes a monotonic, steep decline followed by weak irregular 
oscillations with no clear decay envelope.
In other words, when the disordered quantum system has a large enough Hilbert 
space, we find that the erasure of the initial state information 
occurs mostly at the beginning of the time evolution.

This motivates us to depart from the usual approach of characterizing the 
decaying envelope, and instead focus on the monotonic decline of $R(t)$ at the 
beginning.
To this end, we impose an (arbitrary) threshold $R_0$ on $R(t)$ and
define an ``effective decoherence time'' $T$ as the moment when $R(t)$ drops 
below $R_0$ for the first time,
\begin{equation}\label{eq:T}
T=\min\{t: R(t)<R_0\}.
\end{equation}
Figure~\ref{fig:ising-rp-dynamics}(c) illustrates the thresholding procedure 
using $R_0=0.5$ on the $N=8$ data as an example.
Changing the threshold $R_0$ will change the absolute scale for $T$ without 
affecting any of its qualitative behavior (which we have checked explicitly).

\begin{figure}[]
\centering
\includegraphics{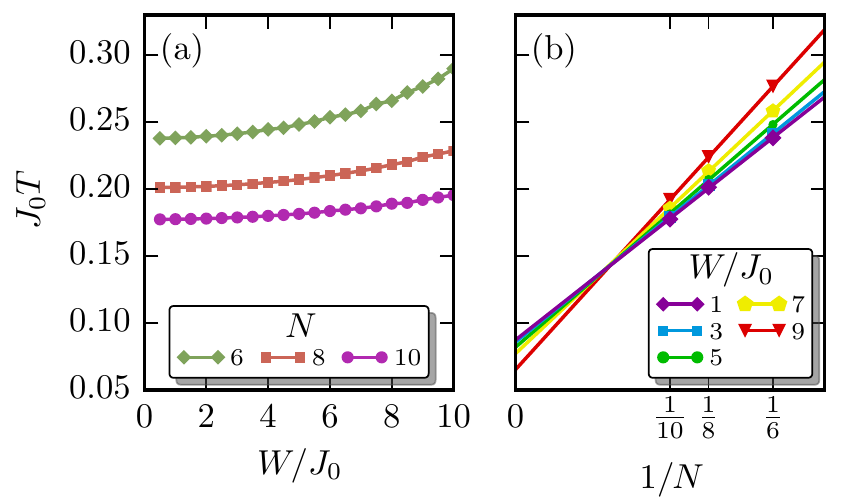}%
\caption{\label{fig:ising-T-pristine}
Dependence of the time scale $T$ extracted from the disorder-averaged return 
probability $R(t)$ on (a) the disorder strength $W$ and (b) the system size 
$N$ of the Ising model, with threshold $R_0=0.75$.
The lines in (a) are a guide to the eye, while those in (b) are linear fits of 
the data points grouped by $W$.
}
\end{figure}

The quantity $T$ defined above characterizes the time needed for the isolated 
system to relax from the initial state, and it may serve as a characteristic 
time scale for such intrinsic decoherence.
In Fig.~\ref{fig:ising-T-pristine}, we examine the dependence of $T$ on the 
disorder strength $W$ for various system sizes $N$.
We observe that $T$ slightly increases as disorder $W$ increases, consistent 
with the intuition that a strong disorder impedes thermalization and induces 
localization.
However, we also find that $T$ suffers from a severe system-size dependence, 
dropping sharply as the number of qubits $N$ increases.
The size dependence completely eclipses the variation of $T$ as the disorder 
strength $W$ changes.
From the linear extrapolation in Fig.~\ref{fig:ising-T-pristine}(b), even the 
qualitative trend of the $W$ dependence of $T$ may shift from increasing to 
decreasing when the system size is large enough.
It therefore appears that although $T$ seems to be a reasonable quantity, its 
use as an effective system decoherence time is fraught with the pitfalls of 
severe finite-size effects.

Such an undesirable behavior of $T$ is rooted in its definition based on 
thresholding the return probability.
For a many-qubit system, the return probability has a strong system-size 
dependence from the dilution effect in the exponentially large Hilbert space. 
As the system size $N$ increases, the fixed threshold $R_0$ on the 
disorder-averaged return probability becomes effectively easier to reach.
Thus, the effective decoherence time $T$ for the whole system turns out to be 
a poor definition or characterization for intrinsic decoherence because of 
uncontrolled finite-size effects, although it may very well work for a system 
with finite and fixed number of qubits.

To get around the system-size dependence, we now introduce the ``per-site'' 
return probability for a system of $N$ qubits by taking the $N$-th root of the 
usual return probability,
and define its disorder average as
\begin{equation}
R_\text{site}(t)=
\average{\big|\langle\psi(t)|\psi(0)\rangle\big|^{2/N}}.
\end{equation}
Here, the $N$-th root effectively folds the exponentially large Hilbert space 
of $N$ qubits so that $R_\text{site}(t)$ behaves like the effective fidelity 
of a single qubit.
This rescaling can also be understood in a spirit similar to the per-site 
error in density-matrix-renormalization-group calculations~\cite{Schollwock11:DMRG}.
Using the rescaled return probability, we define the intrinsic decoherence 
time
\begin{equation}\label{eq:T-site}
T_\text{site}=\min\{t: R_\text{site}(t)<R_0\}
\end{equation}
as the time when $R_\text{site}(t)$ drops below a given 
threshold $R_0$ for the first time.

\begin{figure}[]
\centering
\includegraphics{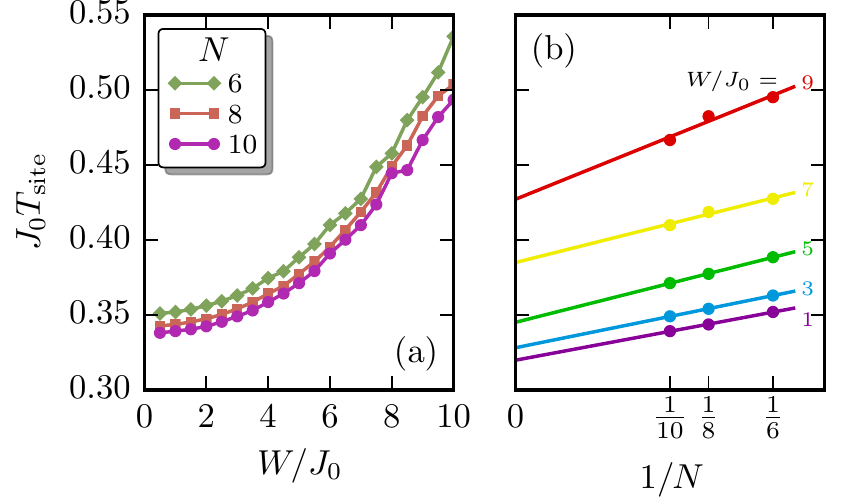}%
\caption{\label{fig:ising-T-site}
Dependence of the intrinsic decoherence time $T_\text{site}$ extracted from 
the disorder-averaged per-site return probability $R_\text{site}(t)$
on (a) the disorder strength $W$ and (b) the system size $N$ of the Ising 
model, with threshold $R_0=0.9$.
The lines in (a) are a guide to the eye, while those in (b) are linear fits of 
the data points grouped by the disorder strength $W$.
}
\end{figure}

In Fig.~\ref{fig:ising-T-site}(a) we find that the intrinsic decoherence time 
$T_\text{site}$ increases with the disorder strength $W$
because relaxation or thermalization of the many-body system is slowing down 
with increasing disorder.
Further, the data for different system sizes almost collapse into a single 
curve. The residual system-size dependence is more clearly visible in 
Fig.~\ref{fig:ising-T-site}(b).
We find that $T_\text{site}$ has an approximately linear dependence on the 
inverse system size $1/N$, and the slope of this dependence does not change 
significantly as disorder $W$ changes.
It should be noted that varying the threshold $R_0$ affects the precise value 
of $T_\text{site}$ but does not introduce any qualitative change to its 
dependence on disorder or system size.
Hence, the intrinsic decoherence time $T_\text{site}$ defined in 
Eq.~\eqref{eq:T-site} provides a system-size-insensitive universal
measure of the rate at which an isolated quantum system loses its memory of 
the collective initial state.

As a side note, the decoherence time $T_\text{site}$ can also be viewed as a 
characterization of the ``speed'' of the disorder-averaged return probability 
dynamics, and it may be subject to a lower bound imposed by the counterpart of 
the ``quantum speed limit''~\cite{Margolus98,Levitin09} in disordered systems.
Establishing a precise connection between the quantum speed limit and the 
decoherence time remains an interesting open question for the future.

Despite the rescaling, $T_\text{site}$ is still defined from a global quantity 
that cannot be measured locally.
It is thus instructive to compare the behavior of $T_\text{site}$ with the 
corresponding decay time of local observables.
As a specific example, we compute the dynamics of the normalized Hamming 
distance~\cite{Hauke15:Hamming,Smith15:MBL}.
For the N\'eel initial state that we consider, the disorder average of the 
Hamming distance is simply given by
\begin{equation}
\mathcal{D}(t)=\frac{1}{2}-\frac{1}{2N}\sum_j^N(-1)^j
\average{\langle\psi(t)|\sigma_j^z|\psi(t)\rangle}.
\end{equation}

\begin{figure}[]
\centering
\includegraphics{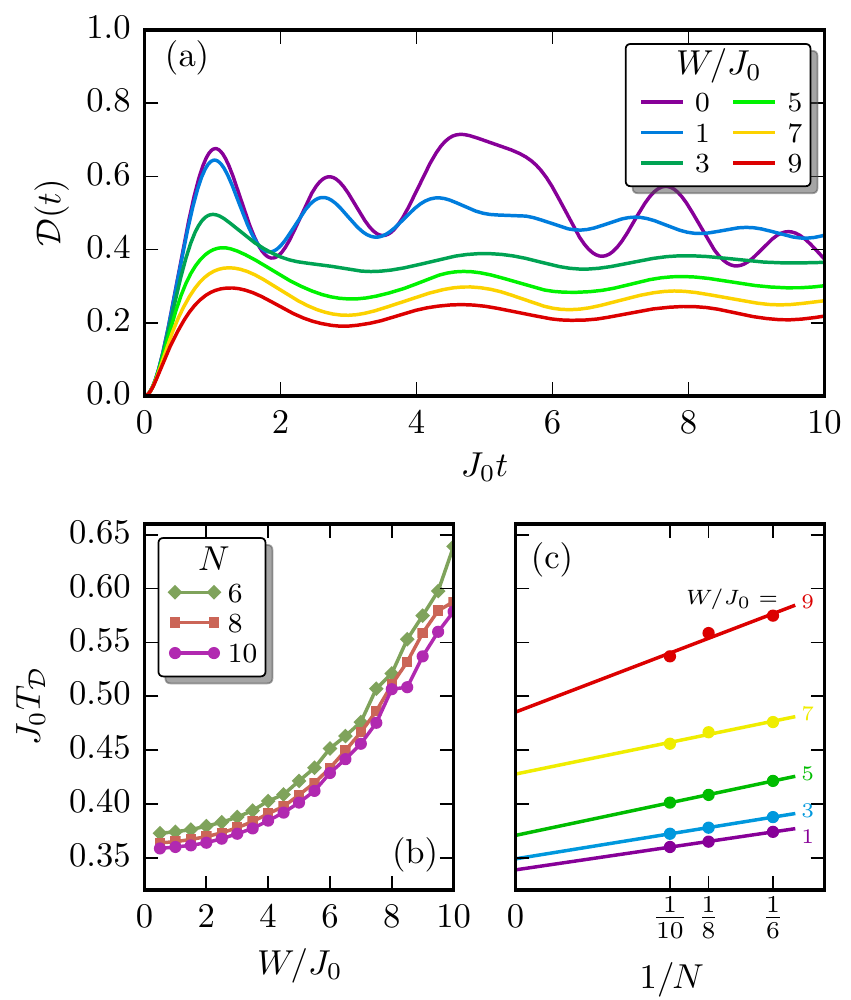}%
\caption{\label{fig:ising-T-hamming}
(a) Dynamics of the Hamming distance from the initial N\'eel state of the 
Ising spins.
Dependence of the intrinsic decoherence time $T_\mathcal{D}$ extracted from 
the disorder-averaged Hamming distance $\mathcal{D}(t)$
on (b) the disorder strength $W$ and (c) the system size $N$, with threshold 
$\mathcal{D}_0=0.25$.
The lines in (b) are a guide to the eye, while those in (c) are linear fits of 
the data points grouped by the disorder strength $W$.
}
\end{figure}

Figure~\ref{fig:ising-T-hamming}(a) shows the time dependence of $\mathcal{D}(t)$ 
for various strengths of disorder $W$.
We find that the dynamics of $\mathcal{D}(t)$ consists of an initial stage of 
monotonic increase followed by weak but irregular oscillations.
Similar to the handling of the return probability $R(t)$, we can define 
another measure of the intrinsic decoherence by thresholding $\mathcal{D}(t)$:
\begin{equation}
T_\mathcal{D}=\min\{t: \mathcal{D}(t)>\mathcal{D}_0\}.
\end{equation}
Although the threshold $\mathcal{D}_0$ here affects the precise value of 
$T_\mathcal{D}$, varying it does not introduce any qualitative change to the 
monotonic dependence of $T_\mathcal{D}$ on disorder.
It should be noted that setting too high a threshold $\mathcal{D}_0$ may send 
$T_\mathcal{D}$ to infinity, when the Hamming distance $\mathcal{D}(t)$ never 
reaches $\mathcal{D}_0$ for sufficiently strong disorder.

Figures~\ref{fig:ising-T-hamming}(b) and~\ref{fig:ising-T-hamming}(c) show the 
dependence of this 
alternative measure on the disorder strength as well as the system size.
Similar to $T_\text{site}$, here we find that $T_\mathcal{D}$ also increases 
with the disorder strength $W$ and is largely insensitive to system size.
This consistency between the decoherence time scales extracted from the 
per-site return probability and from local observables lends further support 
to our procedure of quantifying the intrinsic decoherence time.

In the rest of this paper, we provide further characterizations of the 
intrinsic decoherence time in three additional models, namely the Heisenberg 
model, the interacting Anderson model, and the interacting Aubry-Andr\'e model.
Throughout we use the definition of intrinsic decoherence time $T_\text{site}$ 
as defined and discussed above on a per-site basis in order to eliminate 
finite-size effects.

\section{Heisenberg model} 
\label{sec:Heisenberg}

\begin{figure}[t]
\centering
\includegraphics[width=0.9\linewidth]{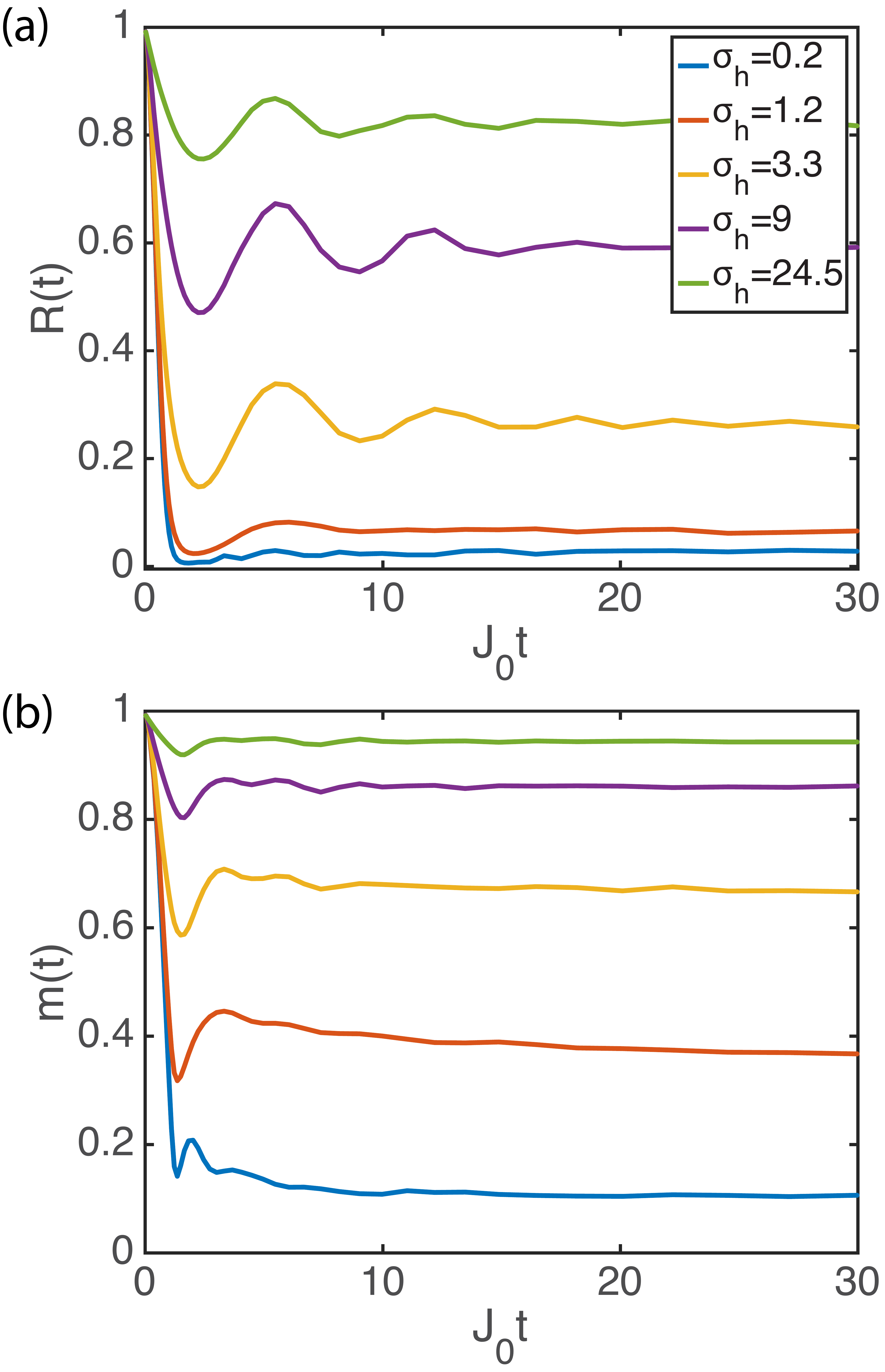}%
\caption{
Dynamics of (a) the return probability and (b) the local magnetization for the 
Heisenberg model with $N=12$ spins, $\sigma_J=0.1J_0$, and different disorder 
strengths $\sigma_h$.
(a) and (b) share the same legend.
} 
\label{fig:RPt-Mt}
\end{figure}

In this section, we study the intrinsic decoherence time in the Heisenberg 
model.
We consider a one-dimensional array of $N$ quantum dots, each hosting one 
localized electron~\cite{Hanson2007Spins,Loss1998Quantum}.
This model is the appropriate description for the so-called exchange-gate 
architectures in semiconductor spin quantum computation,
with the neighboring dots coupled through an exchange coupling arising from 
the combination of interdot Coulomb interaction and the single-particle 
interdot wave-function overlap~\cite{Stafford94,Hu00,Scarola05}.
The collective dynamics of the coupled spin qubits under 
a local magnetic field is described by the following Heisenberg 
Hamiltonian~\cite{Barnes2016Noise}:
\begin{equation}
\label{HeisenbergM}
H=\sum_{k=1}^NJ_k \mathbf{S}_k\cdot \mathbf{S}_{k+1}+\sum_{k=1}^Nh_kS^z_k.
\end{equation}
Here, the local magnetic field $h_k$ and the exchange coupling $J_k$ are 
independent random variables drawn from the normal distributions 
$\mathcal{N}(0,\sigma_h^2)$ and $\mathcal{N}(J_0,\sigma_J^2)$, respectively.
The standard deviations $\sigma_h$ and $\sigma_J$ describe the strengths of 
Overhauser and charge noise~\cite{Dial2013Charge}, respectively, with the 
Overhauser noise being a measure of the background fluctuations in the local 
magnetic field at the qubits (which could arise, for example, from the very 
slow nuclear fluctuations whose dynamics is being ignored here).
The total spin $S^{z}=\sum_{k}S_{k}^{z}$ is conserved, and we focus only on 
the $S^{z}=0$ sector to simplify the calculation.
We truncate the charge noise distribution to the region $J_k>0$ since the 
quantum dot exchange couplings are typically positive~\cite{Barnes2016Noise}.

\begin{figure}[t]
\centering
\includegraphics[angle=0,width=\linewidth]{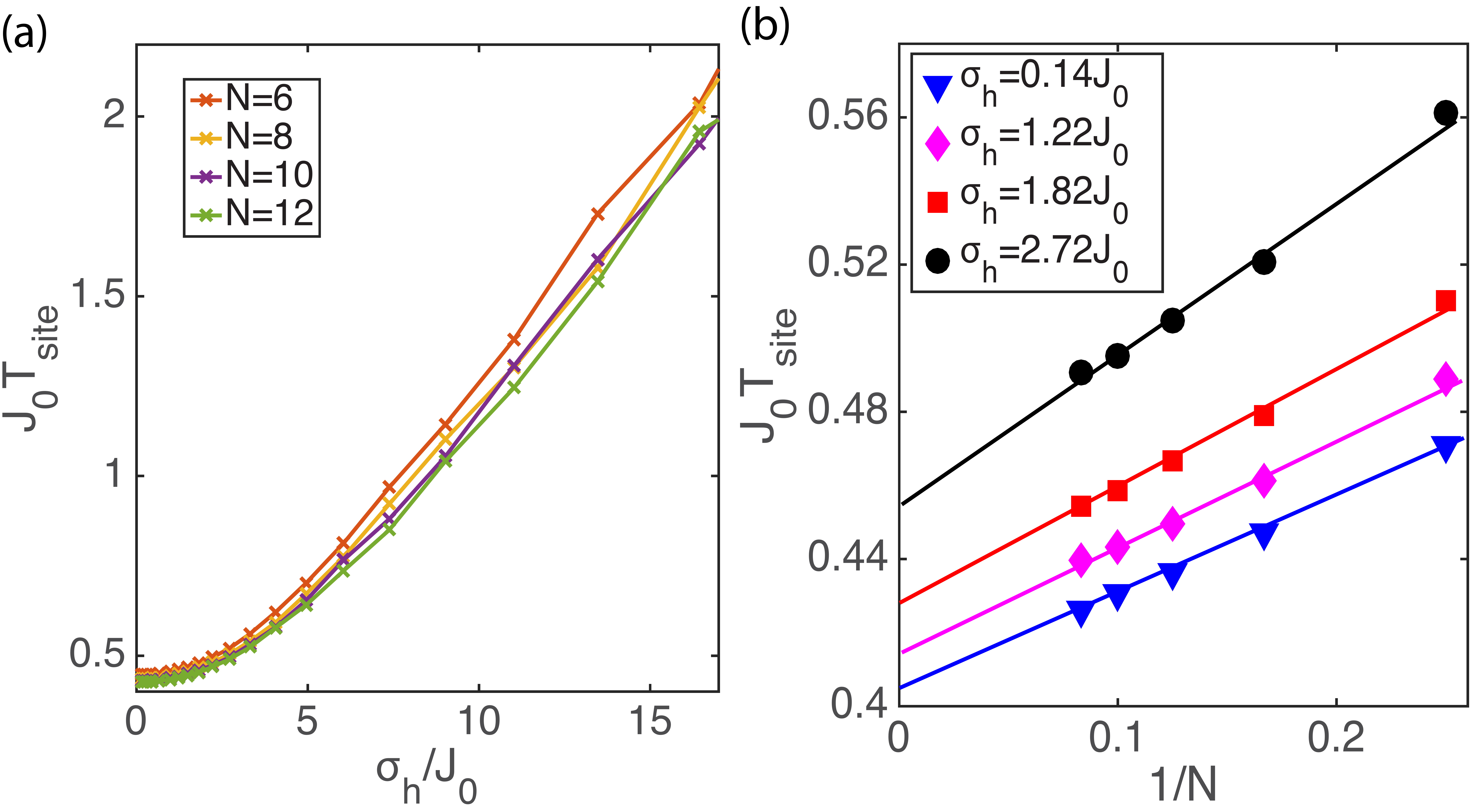}%
\caption{
(a) Intrinsic decoherence time
determined from the per-site return probability  
[defined in Eq.\ref{eq:T-site}], as a function of increasing Overhauser noise 
strength $\sigma_h$, for the Heisenberg model with different system sizes.
(b) System-size dependence of 
$T_{\text{site}}$ for different $\sigma_h$. The lines in (b) are linear fits 
of the data. Here, we have fixed $\sigma_J=0.1J_0$ and the threshold value is 
$R_0=0.95$.}
\label{fig:TpersiteVSsigmah}
\end{figure}

Similar to the previous section on trapped-ion qubits, here we also consider 
the antiferromagnetic N\'{e}el initial state 
$\ket{\psi(0)}=\ket{\uparrow\downarrow\cdots\uparrow\downarrow}$ and
track the time evolution
$|\psi(t)\rangle=e^{-iHt}|\psi(0)\rangle$ computed from exact diagonalization.
We calculate the disorder-averaged
return probability $R(t)$ [see Eq.~\eqref{eq:R}] and the local magnetization 
$m(t)$ commonly studied in the many-body-localization 
literature~\cite{Nandkishore15:MBLReview,Pal10:MBL,Imbrie14:Proof},
\begin{equation}
\label{LocalMag}
m(t)=\left\llbracket
\frac{1}{N}\sum_k \Big|\langle\psi(t)| \sigma_k^z |\psi(t)\rangle\Big|
\right\rrbracket,
\end{equation}
where $\sigma_k^z=2S_k^z$ is the usual Pauli-Z matrix.
In this section, the number of random realizations used ranges from $10^4$ for 
$L=4$ to $10^3$ for $L=12$.

In Fig.~\ref{fig:RPt-Mt}, we plot the time dependence of the return probability 
[Fig.~\ref{fig:RPt-Mt}a] and the local magnetization [Fig.~\ref{fig:RPt-Mt}b] 
for $N=12$. We find that for small $\sigma_h$, both the return probability and 
the local magnetization decay quickly, while for large $\sigma_h$, they first 
decrease a little bit and then remain stable at some finite value. This can be 
understood from the many-body localization-delocalization 
perspective~\cite{Nandkishore15:MBLReview}.
For small $\sigma_h$, the system is in an extended delocalized phase and 
thermalizes quickly, leading to the fast decay of $R(t)$ and $m(t)$.
In the thermodynamic limit $L\rightarrow\infty$, both $R(t)$ and $m(t)$ 
eventually vanish at large $t$.
On the other hand, by increasing $\sigma_h$, the system will eventually go 
through a many-body-localization transition into a localized phase at some 
large critical disorder. In the localized region, the dynamics is drastically 
suppressed and certain local information is preserved even in the infinite 
time limit. Thus, in this localized large-disorder region, $R(t)$ and $m(t)$ 
first decay a little bit (mainly due to the wave-packet expansion within the 
localization length) and then remain stable at a finite value for a finite 
system size because of memory retention in the many-body-localized state.
In this large-disorder localized phase the effective decoherence time then 
becomes infinite in the thermodynamic limit as the system simply fails to 
thermalize.

\begin{figure}[t]
\centering
\includegraphics[angle=0,width=\linewidth]{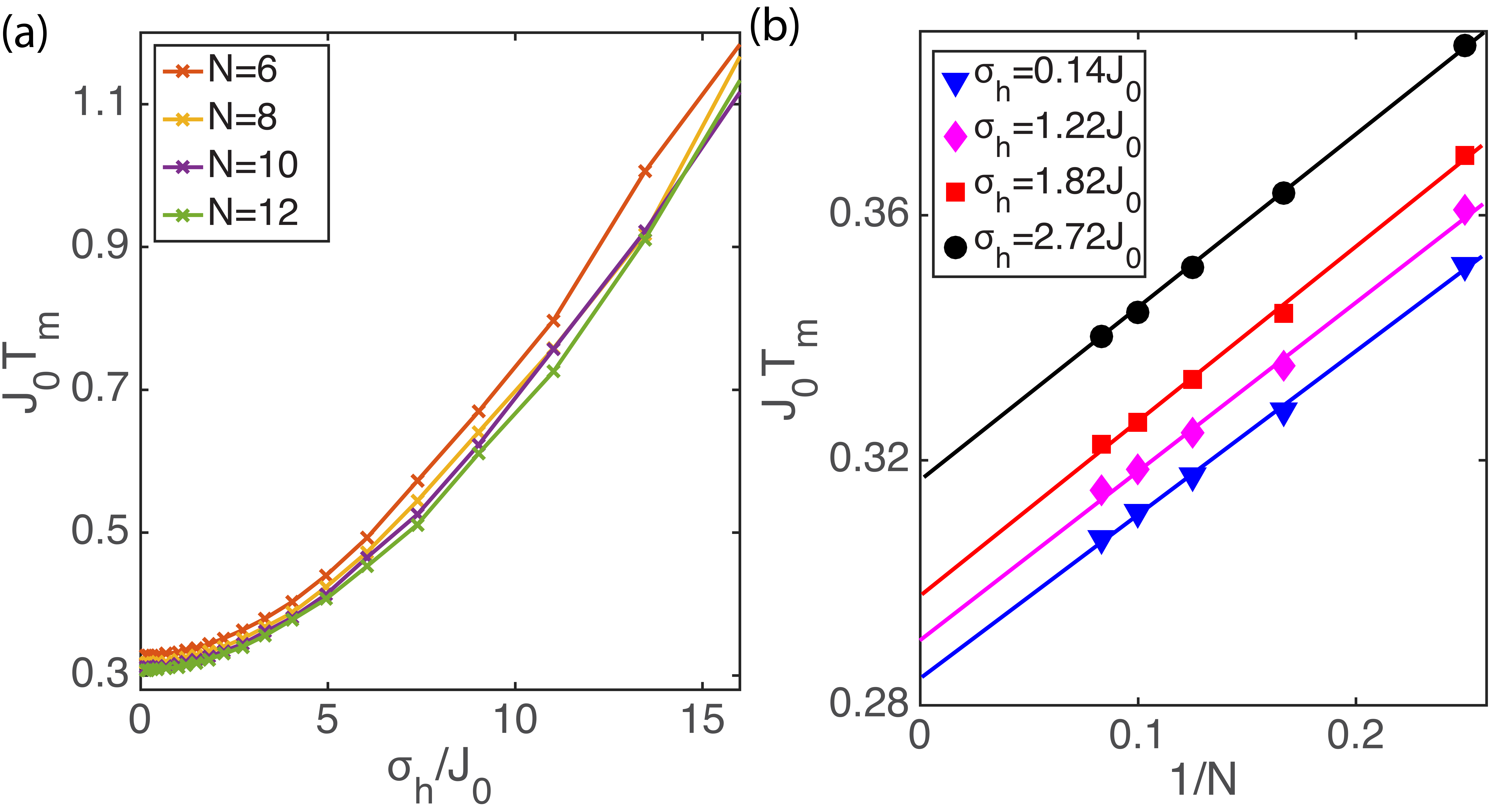}%
\caption{
(a) Intrinsic decoherence time determined from the local magnetization $m(t)$, 
as a function of increasing Overhauser noise strength $\sigma_h$, for the 
Heisenberg model with 
different system sizes. (b) System-size dependence of $T_m$ for different 
$\sigma_h$. The lines are linear fits of the data. Here, the threshold value 
is chosen to be $m_0=0.9$ and we have fixed $\sigma_J=0.1J_0$.
}
\label{fig:TmagVSsigmah}
\end{figure}

We now turn to the study of the intrinsic decoherence time. We first focus on 
the case of return probability. As discussed in the previous section, we can 
extract the intrinsic decoherence time $T_{\text{site}}$ 
[see Eq.~\eqref{eq:T-site}] from the per-site 
return probability. In Fig. \ref{fig:TpersiteVSsigmah}(a), we find that the 
calculated intrinsic decoherence time increases with increasing Overhauser 
noise strength $\sigma_h$, as expected from the localization physics. Figure
\ref{fig:TpersiteVSsigmah}(b) shows the system-size dependence of the 
extracted $T_{\text{site}}$. Similar to the transverse-field Ising model, an 
approximately linear dependence of $T_{\text{site}}$ on the inverse system 
size $1/N$ is obtained for each fixed $\sigma_h$. Moreover, the slope of the 
linear dependence does not vary much as $\sigma_h$ varies.
This strong suppression of finite-size effect again explicitly shows the advantage 
of defining the intrinsic decoherence time based on per-site return probability.

As shown in Fig.~\ref{fig:RPt-Mt}(b), we have also computed the dynamics of 
the local magnetization $m(t)$ for different disorder strengths $\sigma_h$.
Similar to $T_\text{site}$ defined in the previous section, we can construct
another measure of the intrinsic decoherence by thresholding $m(t)$:
\begin{eqnarray}
T_m=\min\{t: m(t)<m_0\}. \label{Defin-Tm}
\end{eqnarray}
Figure~\ref{fig:TmagVSsigmah}(a) shows the dependence of $T_m$ on the Overhauser 
strength $\sigma_h$. We find again that the such-defined intrinsic decoherence 
time increases with increasing $\sigma_h$.
In addition, the data for different system sizes collapse approximately into a 
single curve, manifesting an apparent scaling behavior.
The residual system-size dependence of $T_m$ 
is shown in Fig. \ref{fig:TmagVSsigmah}(a).  Similar to the case of 
$T_{\text{site}}$, we find that $T_m$ has a linear dependence on $1/N$ and the 
slope of the dependence does not change significantly as $\sigma_h$ changes.

\section{Interacting Anderson model} 
\label{sec:Anderson}

\begin{figure}[t]
\centering
\includegraphics[width=\linewidth]{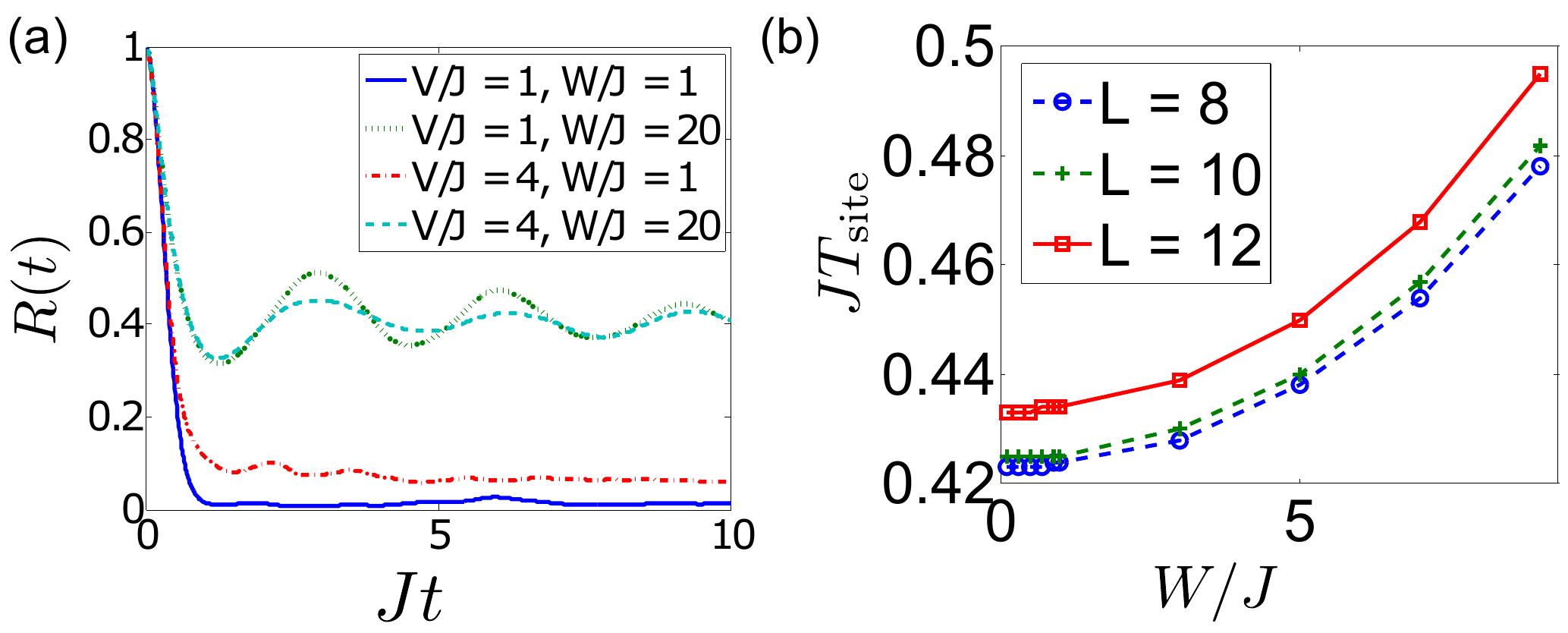}%
\caption{
Disorder-averaged dynamics of the return probability and the corresponding 
decoherence time for the interacting Anderson model. We chose initial states with fermions occupying random lattice sites. 
(a) The dynamical evolution of the return probability $R(t)$  for 
different interaction $V/J$ and disorder strengths $W/J$.
We average over $10^3$ disorder realizations with one random initial 
state for each realization.
(b) The intrinsic decoherence time $T_{\rm site}$ 
[Eq.~\eqref{eq:T-site}] with the threshold $R_0=0.9$.
In (b), the interaction strength is fixed to be $V/J =1$. Using different 
interaction strengths does not change the qualitative features shown here.
}
\label{fig:RTCombine}
\end{figure}

\begin{figure}[t]
\centering
\includegraphics[width=\linewidth]{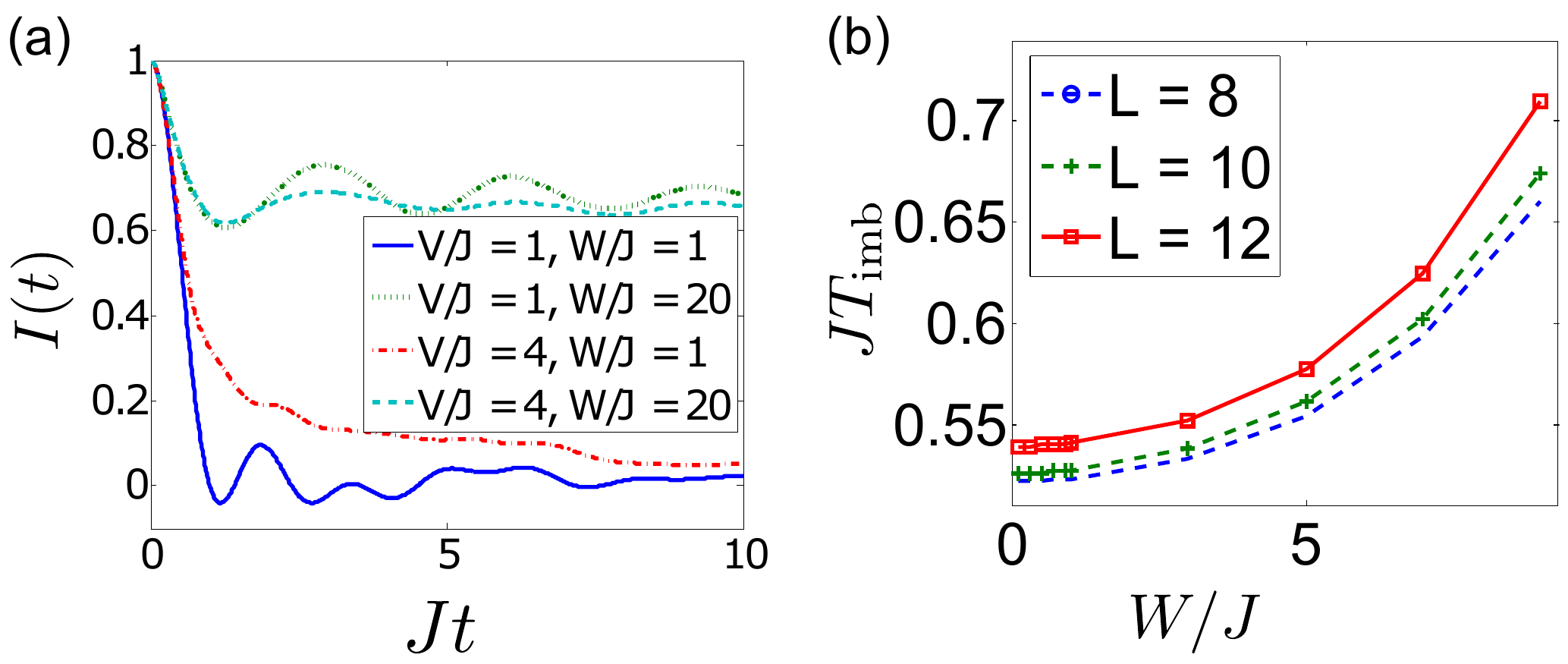}%
\caption{
Disorder-averaged dynamics of the number imbalance
and the corresponding relaxation time for the interacting 
Anderson model. We chose initial states with fermions occupying random sites. 
(a) The dynamics of the disorder-averaged number imbalance $I(t)$ for 
different interaction $V/J$ and disorder strengths $W/J$. We average over 
$10^3$ disorder realizations with one random initial state in each 
realization. (b) shows the decoherence time $T_{\rm imb}$ as determined by 
the time $I(t)$ decays to the threshold $I_0=0.5$.
In (b) the interaction strength is fixed to be $V/J$ =1.
}
\label{fig:ImbCombine}
\end{figure}

As another example, in this section we study intrinsic decoherence in the 
interacting Anderson model:
\begin{align}
 H &= \sum_{i} -J [c_i^\dag c_{i+1}+c_{i+1}^\dag c_i] +  h _i c_i^\dag c_i \nonumber  \\
& + V\sum_{i} c_i^\dag c_{i+1} ^\dag c_{i+1} c_i.
\label{eq:HamAnderson} 
\end{align}
Here $J$ is the tunneling strength and $V$ is the interaction strength between 
nearest neighbors. The random potential $h_i$ is drawn uniformly from 
$[-W/2, W/2]$. 
We focus on half filling for the interacting Anderson model. This model has 
been widely studied in the context of many-body localization~\cite{Basko06:MBL,
Oganesyan07:MBL,Bardarson12:EEGrowth,Bauer13:AreaLaw,Iyer15:MBL,Deng15:MBL,Deng16:MBL}.
With the interaction strength $V/J = 2$, the model reduces to the Heisenberg model in Eq.~\eqref{HeisenbergM} with $\sigma_J = 0$.  
 
Using exact diagonalization, we calculate the relaxation dynamics of the 
disorder-averaged return probability 
and the number imbalance~\cite{Schreiber15:MBL} relaxation for initial product 
states with particles occupying random lattice sites.
Note that this choice is different from the N\'eel states used for spin models. 
The average number imbalance is defined by
\begin{equation}
I (t) = \average{\frac{ N_{1} (t)- N_{0}(t)}{N_1(t) + N_0(t)}},  
\end{equation}
with $N_1$ ($N_0$) referring to the particle number in the initially occupied 
(unoccupied) sites~\cite{Li16:bubbleMBL}
and the double brackets $\average{\cdot}$ now denoting an average over both 
disorder realizations and initial state configurations.
The dynamical evolution of the return probability and the corresponding 
decoherence time are shown in Fig.~\ref{fig:RTCombine}. 
The number imbalance dynamics and its decoherence time $T_\text{imb}$ are shown in 
Fig.~\ref{fig:ImbCombine}, with $T_\text{imb}$ defined by thresholding $I(t)$ 
in a similar fashion to Eq.~\eqref{Defin-Tm}.

When the disorder $W$ is weak, both the return probability and the number 
imbalance decay quickly to almost zero, indicating a fast memory loss in the 
thermal phase.
The intrinsic decoherence times determined from the return probability and the 
number imbalance qualitatively agree with each other.
Both of them increase monotonically as we increase the disorder 
strength. In contrast, when the system is in a many-body localized phase at 
strong disorder, the number imbalance and the per-site return probability 
remain finite even in the long time limit.
The corresponding monotonic increase of the decoherence time with increasing 
disorder strength provides a quantitative characterization of the memory 
retention protected by localization.

Compared with the two spin models, we find that the interacting Anderson model 
has a slightly stronger finite-size effect. This is manifested in the scaling 
of both $T_\text{site}$ determined from the return probability and 
$T_\text{imb}$ determined from the number imbalance.

\begin{figure}[t]
\centering
\includegraphics[width=\linewidth]{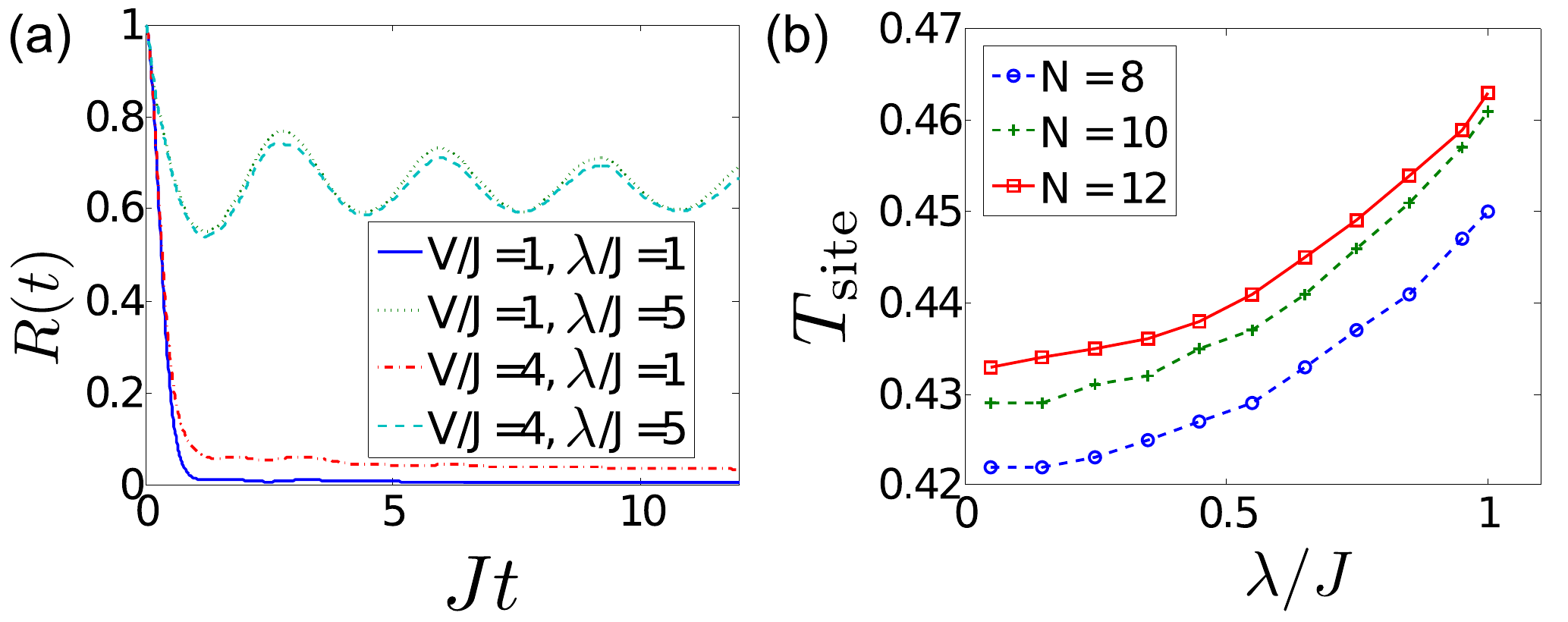}%
\caption{
Dynamics of the return probability and the corresponding 
decoherence time for the interacting Aubry-Andr\'e  model. In the initial state, fermions occupy random lattice sites. 
(a) The dynamical evolution of the return probability $R(t)$  for 
different interaction $V/J$ and incommensurate lattice potential strengths
$\lambda/J$.
We average over $10^3$ different $\phi$ realizations [Eq.~\eqref{eq:AA}].
(b) The intrinsic decoherence time $T_{\rm site}$ 
[Eq.~\eqref{eq:T-site}] with the threshold $R_0=0.9$.
In (b), the interaction strength is fixed to be $V/J =1$. 
}
\label{fig:AAQRT}
\end{figure}

\begin{figure}[t]
\centering
\includegraphics[width=\linewidth]{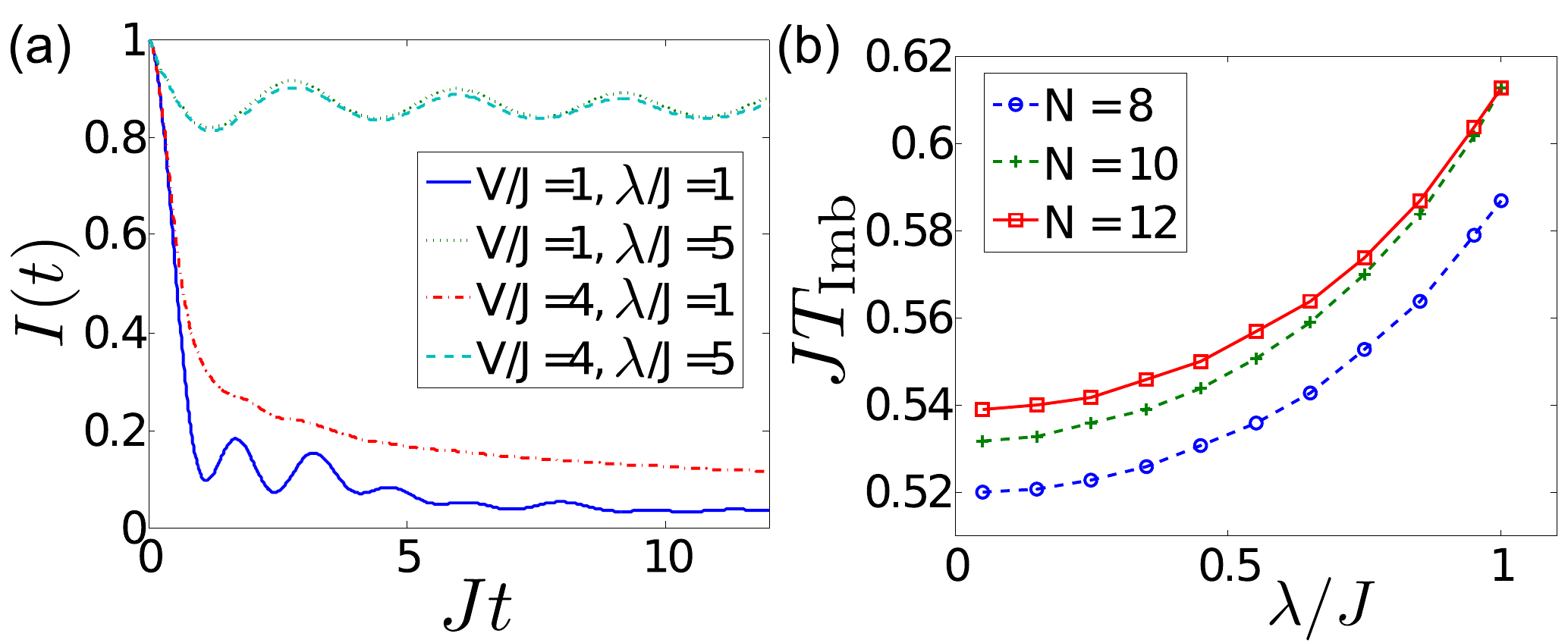}%
\caption{
The number imbalance dynamics 
and the corresponding relaxation time for the interacting 
Aubry-Andr\'e model.
(a) The dynamics of the number imbalance $I(t)$ with varying 
interaction $V/J$ and incommensurate potential strengths $\lambda/J$.
We average over $10^3$ different $\phi$-realizations with one random initial 
state (fermions occupying random sites) in each 
realization. (b) The decoherence time $T_{\rm imb}$ as determined by 
the time $I(t)$ decays to the threshold $I_0=0.5$.
In (b) the interaction strength is fixed to be $V/J$ =1.
}
\label{fig:AAimb}
\end{figure} 

\section{Interacting Aubry-Andr\'e model} 
\label{sec:AA} 
In this section, we study the decoherence for the interacting
Aubry-Andr\'e model~\cite{Aubry80:AA,Azbel64:AA,Harper55:AA,Grempel82:AA,Iyer15:MBL,Li15:MBL,Modak15:MBL}.
This model can be defined by replacing the random potential $h_j$ in 
Eq.~\eqref{eq:HamAnderson} with an incommensurate lattice potential
\begin{equation} 
h_j = 2\lambda \cos (2\pi Q j + \phi),
\label{eq:AA}
\end{equation} 
where $Q$ is an irrational number, here chosen to be the golden ratio. 

In parallel with the interacting Anderson model, we have calculated the 
dynamics of the return probability and the number imbalance for the 
Aubry-Andr\'e model following initial states with fermions occupying 
random sites. As shown in Figs.~\ref{fig:AAQRT} and~\ref{fig:AAimb},
the numerical data are qualitatively similar to those of the Anderson model, 
except with a slightly stronger finite-size effect.
We still find a monotonic dependence of the decoherence times on the disorder 
strength.

\section{Conclusion} 
\label{sec:conclusion}
In this paper we have studied the intrinsic decoherence in the time evolution 
of an isolated quantum system due to disorder.
We analyze the scaling of the return probability as a function of system size 
and propose a quantitative, system-size-insensitive measure for the erasure of 
the initial state memory. 
We call this measure the intrinsic decoherence time of the isolated quantum 
system.
Using four different models, we characterize the dependence of the intrinsic 
decoherence time on both the system size and the disorder strength, and we 
compare it with other time scales in the relaxation of local observables.
We find that our definition of the intrinsic decoherence time consistently
captures the time scale associated with the initial state memory retention, and 
it increases monotonically as the system becomes more disordered.
Our results introduce quantitative measures on how fast an isolated 
disordered system forgets its initial state information and may provide
useful guidance for future experiments. 
In particular, experiments in cold atomic gases and ion traps could mimic our 
results for the Anderson model and the Ising model, respectively, whereas our 
Heisenberg model results can be compared eventually with multiqubit
quantum-dot-based semiconductor spin qubit systems.

We emphasize that the intrinsic decoherence time studied theoretically in our 
work is a physical measure characterizing the loss of memory in disordered 
quantum interacting systems, which should be directly accessible in 
experiments through the measurement of time-dependent magnetization (or number 
imbalance).  Our work can thus be used for a quantitative study of many-body 
localization and decoherence phenomena of wide interest.  We also emphasize 
that the precise value of the intrinsic decoherence time would depend not only 
on the system Hamiltonian, but also on the starting initial state, which 
must be a noneigenstate of the Hamiltonian.  We have used reasonable physical 
initial states (e.g., N\'eel states for our Ising and Heisenberg studies) because 
these are easy to prepare in the laboratory, but the qualitative behavior of 
the intrinsic decoherence time should not depend on the precise choice of the 
initial state (or the precise thresholding procedure) as long as the system 
starts in a generic unentangled product state.

\section*{Acknowledgment} 
This work is supported by JQI-NSF-PFC and LPS-MPO-CMTC.
We acknowledge the University of Maryland supercomputing resources
(http://www.it.umd.edu/hpcc) made available in conducting the research 
reported in this paper.

\end{document}